\documentstyle[12pt]{article}
\newcommand\be{\begin{equation}}
\newcommand\ee{\end{equation}}
\newcommand\bea{\begin{eqnarray}}
\newcommand\eea{\end{eqnarray}}
\newcommand\ket[1]{|#1\rangle}

\newcommand\braket[2]{\langle #1|#2\rangle}
\newcommand{\fatalpha}{{\bf \alpha \kern -0.44em \alpha}}
\newcommand{\fatsigma}{{\bf \sigma \kern -0.54em \sigma}}
\newcommand{\tpchi}{{\bf \chi \kern -0.35em \chi}}
\newcommand{\llambda}{{\bf \lambda \kern -0.45em \lambda}}

\bibliography{plain}
\title{\bf Perfect transfer of $m$-qubit GHZ states}\vspace{20mm}
\author{ M. A. Jafarizadeh$^{a,b,c,d}$
 \thanks{E-mail:jafarizadeh@tabrizu.ac.ir},
 R. Sufiani$^{a,c}$, S. F. Taghavi$^{a}$ and E. Barati$^{a}$
 \thanks{E-mail:sofiani@tabrizu.ac.ir}
 \\ $^a${\small Department of Theoretical Physics and Astrophysics,
University of Tabriz, Tabriz 51664, Iran.} \\ $^b${\small Center
of excellence for photonic, University of Tabriz, Tabriz 51664,
Iran.}
\\ $^c${\small Institute for Studies in
Theoretical Physics and Mathematics, Tehran 19395-1795, Iran.} \\
$^d${\small Research Institute for Fundamental Sciences, Tabriz
51664, Iran.}} \pagebreak

\vspace{20mm}
\begin{document}
\maketitle \vspace{15mm}
\newpage
\begin{abstract}
By using some techniques such as spectral distribution and
stratification associated with the graphs, employed in
\cite{PST,pstd} for the purpose of Perfect state transfer (PST) of
a single qubit over antipodes of distance-regular spin networks
and PST of a $d$-level quantum state over antipodes of
pseudo-distance regular networks, PST of an $m$-qubit GHZ state is
investigated. To do so, we employ the particular distance-regular
networks (called Johnson networks) $J(2m,m)$ to transfer an
$m$-qubit GHZ state initially prepared in an arbitrary node of the
network (called the reference node) to
the corresponding antipode, perfectly.\\


{\bf PACs Index: 01.55.+b, 02.10.Yn }
\end{abstract}

\vspace{70mm}
\newpage
\section{Introduction}
In the interior of quantum computers good communication between
different parts of the system is essential. The need is thus to
transfer quantum states and generate entanglement between
different regions contained within the system. In quantum
information processing (QIP) protocols, quantum spin systems
\cite{1',3',10'} (or quasi-spin \cite{4'}) systems serve as
quantum channels. The idea to use quantum spin chains for short
distance quantum communication was put forward by Bose
\cite{Bose}. He showed that an array of spins (or spin like two
level systems) with isotropic Heisenberg interaction is suitable
for quantum state transfer. In particular, spin chains can be used
as transmission lines for quantum states without the need to have
controllable coupling constants between the qubits or complicated
gating schemes to achieve high transfer fidelity. After the work
of Bose, the use of spin chains \cite{2''}-\cite{spin2} and
harmonic chains \cite{13''} as quantum wires have been proposed.
In the previous work \cite{PST}, the so called distance-regular
graphs have been considered as spin networks (in the sense that
with each vertex of a distance-regular graph a qubit or a spin was
associated ) and perfect state transfer (PST) of a single qubit
state over antipodes of these networks has been investigated,
whereas in Ref.\cite{pstd} the authors have been considered the
PST of a $d$-level quantum state over antipodes of pseudo-distance
regular graphs. In both of these works, a procedure for finding
suitable coupling constants in some particular spin Hamiltonians
has been given so that perfect transfer of a quantum state between
antipodes of the networks can be achieved. The present work
focuses on the PST of the $m$ qubit maximally entangled GHZ states
$\ket{\psi}_{GHZ}=\frac{1}{\sqrt{2}}(\ket{\underbrace{00\ldots
0}_m}+\ket{\underbrace{11\ldots 1}_m})$. To this end, we will
consider the Johnson networks $J(2m,m)$ (which are
distance-regular) as spin networks. Then, we use the algebraic
properties of these networks in order to find suitable coupling
constants in some particular spin Hamiltonians so that perfect
transfer of $m$-qubit GHZ states between antipodes of the networks
can be achieved.

The organization of the paper is as follows: In section 2, we
review some preliminary facts about graphs, their stratifications
, spectral distribution associated with them and underlying
networks derived from symmetric group $S_n$ called also Johnson
networks. Section $3$ is devoted to perfect transfer of $m$-qubit
GHZ states over antipodes of the networks $J(2m,m)$, where a
method for finding suitable coupling constants in particular spin
Hamiltonians so that PST be possible, is given. The paper is ended
with a brief conclusion.
\section{Preliminaries}
In this section we recall some preliminary facts about graphs,
their stratifications, spectral distribution associated with them
and underlying networks derived from symmetric group $S_n$ called
also Johnson networks.
\subsection{Graphs and their stratifications}
A graph is a pair $\Gamma=(V,E)$, where $V$ is a non-empty set
called the vertex set and $E$ is a subset of $\{(x,y):x,y \in
V,x\neq y\}$ called the edge set of the graph. Two vertices $x, y
\in V$ are called adjacent if $(x,y)\in E$, and in that case we
write $x \sim y$. For a graph $\Gamma=(V,E)$, the adjacency matrix
$A$ is defined as
\begin{equation}\label{adj.}
\bigl(A)_{\alpha, \beta}\;=\;\cases{1 & if $\;\alpha\sim \beta$
\cr 0 & \mbox{otherwise}\cr}.
\end{equation}
Conversely, for a non-empty set $V$, a graph structure is uniquely
determined by such a matrix indexed by $V$.

The degree or valency of a vertex $x \in V$ is defined by
\begin{equation}\label{val.}
\kappa(x)=|\{y\in V: y\sim x\}|
\end{equation}
where, $|\cdot|$ denotes the cardinality. The graph is called
regular if the degree of all of the vertices be the same. In this
paper, we will assume that graphs under discussion are regular. A
finite sequence $x_0, x_1, . . . , x_n \in V$ is called a walk of
length $n$ (or of $n$ steps) if $x_{i-1}\sim  x_i$ for all $i= 1,
2, . . . , n$. Let $l^2(V)$ denote the Hilbert space of $C$-valued
square-summable functions on $V$. With each $\beta\in V$ we
associate a vector $\ket{\beta}$ such that the $\beta$-th entry of
it is $1$ and all of the other entries of it are zero. Then
$\{\ket{\beta}: \beta\in V\}$ becomes a complete orthonormal basis
of $l^2(V)$. The adjacency matrix is considered as an operator
acting in $l^2(V)$ in such a way that
\begin{equation}
A\ket{\beta}=\sum_{\alpha\sim \beta}\ket{\alpha}.
\end{equation}

Now, we recall the notion of stratification for a given graph
$\Gamma$. To this end, let $\partial(x,y)$ be the length of the
shortest walk connecting $x$ and $y$ for $x\neq y$. By definition
$\partial(x,x)=0$ for all $x\in V$. The graph becomes a metric
space with the distance function $\partial$. Note that
$\partial(x,y)=1$ if and only if $x\sim y$. We fix a vertex $o \in
V$ as an origin of the graph, called the reference vertex. Then,
the graph $\Gamma$ is stratified into a disjoint union of strata
(with respect to the reference vertex $o$) as
\begin{equation}\label{strat}
V=\bigcup_{i=0}^{\infty}\Gamma_{i}(o),\;\ \Gamma_i(o):=\{\alpha\in
V: \partial(\alpha,o)=i\}
\end{equation}
Note that $\Gamma_i(o)=\emptyset$ may occur for some $i \geq 1$.
In that case we have $\Gamma_i(o)= \Gamma_{i+1}(o)=...=
\emptyset$. With each stratum $\Gamma_i(o)$ we associate a unit
vector in $l^2(V)$ defined by
\begin{equation}\label{unitv}
\ket{\phi_{i}}=\frac{1}{\sqrt{\kappa_i}}\sum_{\alpha\in
\Gamma_{i}(o)}\ket{\alpha},
\end{equation}
where, $\kappa_i=|\Gamma_{i}(o)|$ is called the $i$-th valency of
the graph ($\kappa_i:=|\{\gamma:\partial(o,\gamma)=i
\}|=|\Gamma_{i}(o)|$).

One should notice that, for distance regular graphs, the above stratification is independent of the
choice of reference vertex and the vectors $\ket{\phi_i}, i=0,1,...,d-1$
form an orthonormal basis for the so called Krylov subspace
$K_d(\ket{\phi_0},A)$ defined as
\begin{equation}
K_d(\ket{\phi_0},A) = \mathrm{span}\{\ket{\phi_0},A\ket{\phi_0},
\cdots,A^{d-1}\ket{\phi_0}\}.
\end{equation}
Then it can be shown that \cite{krylov}, the orthonormal basis
$\ket{\phi_i}$ are written as
\be\label{phi}\ket{\phi_i}=P_i(A)\ket{\phi_0},\ee where
$P_i=a_0+a_1A+...+a_iA^i$ is a polynomial of degree $i$ in
indeterminate $A$ (for more details see for example
\cite{krylov,js}).
\subsection{Spectral distribution associated with the graphs}
Now, we recall some preliminary facts about spectral techniques
used in the paper, where more details have been given in Refs.
\cite{js,jss1,js1,jss}.

Actually the spectral analysis of operators is an important issue
in quantum mechanics, operator theory and mathematical physics
\cite{simon, Hislop}. As an example $\mu(dx)=|\psi(x)|^2dx$
($\mu(dp)=|\widetilde{\psi}(p)|^2dp$) is a spectral distribution
which is  assigned to  the position (momentum) operator
$\hat{X}(\hat{P})$. Moreover, in general quasi-distributions are
the assigned spectral distributions of two hermitian non-commuting
operators with a prescribed ordering. For example the Wigner
distribution in phase space is the assigned spectral distribution
for two non-commuting operators $\hat{X}$ (shift operator) and
$\hat{P}$ (momentum operator) with Wyle-ordering among them
\cite{Kim,Hai}. It is well known that, for any pair
$(A,\ket{\phi_0})$ of a matrix $A$ and a vector $\ket{\phi_0}$,
one can assign a measure $\mu$ as follows
\begin{equation}\label{sp1}
\mu(x)=\braket{ \phi_0}{E(x)|\phi_0},
\end{equation}
 where
$E(x)=\sum_i|u_i\rangle\langle u_i|$ is the operator of projection
onto the eigenspace of $A$ corresponding to eigenvalue $x$, i.e.,
\begin{equation}
A=\int x E(x)dx.
\end{equation}
Then, for any polynomial $P(A)$ we have
\begin{equation}\label{sp2}
P(A)=\int P(x)E(x)dx,
\end{equation}
where for discrete spectrum the above integrals are replaced by
summation. Therefore, using the relations (\ref{sp1}) and
(\ref{sp2}), the expectation value of powers of adjacency matrix
$A$ over reference vector $\ket{\phi_0}$ can be written as
\begin{equation}\label{v2}
\braket{\phi_{0}}{A^m|\phi_0}=\int_{R}x^m\mu(dx), \;\;\;\;\
m=0,1,2,....
\end{equation}
Obviously, the relation (\ref{v2}) implies an isomorphism from the
Hilbert space of the stratification onto the closed linear span of
the orthogonal polynomials with respect to the measure $\mu$.

From orthonormality of the unit vectors $\ket{\phi_i}$ given in
Eq.(\ref{unitv}) (with $\ket{\phi_0}$ as unit vector assigned to
the reference node) we have
\begin{equation}\label{ortpo}
\delta_{ij}=\langle\phi_i|\phi_j\rangle=\int_{R}P_i(x)P_j(x)\mu(dx).
\end{equation}
By rescaling $P_k$ as $Q_k=\sqrt{\omega_1\ldots\omega_k}P_k$, the
spectral distribution $\mu$ under question will be characterized
by the property of orthonormal polynomials $\{Q_k\}$ defined
recurrently by
$$ Q_0(x)=1, \;\;\;\;\;\
Q_1(x)=x,$$
\begin{equation}\label{op}
xQ_k(x)=Q_{k+1}(x)+\alpha_{k}Q_k(x)+\omega_kQ_{k-1}(x),\;\;\ k\geq
1.
\end{equation}
The parameters $\alpha_k$ and $\omega_k$ appearing in (\ref{op})
are defined by
\begin{equation}\label{omegal}
\alpha_0=0,\;\;\ \alpha_k=\kappa-b_{k}-c_{k},\;\;\;\;\
\omega_k\equiv\beta^2_k=b_{k-1}c_{k},\;\;\ k=1,...,d,
\end{equation}
where, $\kappa\equiv \kappa_1$ is the degree of the networks and
$b_i$'s and $c_i$'s are the corresponding intersection numbers.
Following Ref. \cite{obata}, we will refer to the parameters
$\alpha_k$ and $\omega_k$ as $QD$ (Quantum Decomposition)
parameters (see Refs. \cite{js,jss1,js1,obata} for more details).
 If such a spectral distribution is unique,
the spectral distribution $\mu$ is determined by the identity
\begin{equation}\label{sti}
G_{\mu}(x)=\int_{R}\frac{\mu(dy)}{x-y}=\frac{1}{x-\alpha_0-\frac{\omega_1}{x-\alpha_1-\frac{\omega_2}
{x-\alpha_2-\frac{\omega_3}{x-\alpha_3-\cdots}}}}=\frac{Q_{d}^{(1)}(x)}{Q_{d+1}(x)}=\sum_{l=0}^{d}
\frac{\gamma_l}{x-x_l},
\end{equation}
where, $x_l$ are the roots of the polynomial $Q_{d+1}(x)$.
$G_{\mu}(x)$ is called the Stieltjes/Hilbert transform of spectral
distribution $\mu$ and polynomials $\{Q_{k}^{(1)}\}$ are defined
recurrently as
$$Q_{0}^{(1)}(x)=1, \;\;\;\;\;\
    Q_{1}^{(1)}(x)=x-\alpha_1,$$
\begin{equation}\label{oq}
xQ_{k}^{(1)}(x)=Q_{k+1}^{(1)}(x)+\alpha_{k+1}Q_{k}^{(1)}(x)+\omega_{k+1}Q_{k-1}^{(1)}(x),\;\;\
k\geq 1,
\end{equation}
respectively. The coefficients $\gamma_l$ appearing in (\ref{sti})
are calculated as
\begin{equation}\label{Gauss}
\gamma_l:=\lim_{x\rightarrow x_l}[(x-x_l)G_{\mu}(x)]
\end{equation}
Now let $G_{\mu}(z)$ is known, then the spectral distribution
$\mu$ can be determined in terms of $x_l, l=1,2,...$ and Gauss
quadrature constants $\gamma_l, l=1,2,... $ as
\begin{equation}\label{m}
\mu=\sum_{l=0}^d \gamma_l\delta(x-x_l)
\end{equation}
(for more details see Refs. \cite{obh,st,tsc,obah}).
\subsection{Underlying networks derived from symmetric group $S_n$}
Let $\lambda=(\lambda_1,...,\lambda_m)$ be a partition of $n$,
i.e., $\lambda_1+...+\lambda_m=n$. We consider the subgroup
$S_m\otimes S_{n-m}$ of $S_n$ with $m\leq [\frac{n}{2}]$. Then we
assume the finite set $M^{\lambda}=\frac{S_n}{S_m\otimes S_{n-m}}$
with $|M^{\lambda}|=\frac{n!}{m!(n-m)!}$ as vertex set. In fact,
$M^{\lambda}$ is the set of $(m-1)$-faces of $(n-1)$-simplex
(recall that, the graph of an $(n-1)$-simplex is the complete
graph with $n$ vertices denoted by $K_n$). If we denote the vertex
$i$ by $m$-tuple $(i_1,i_2,...,i_m)$, then the adjacency matrices
$A_k$, $k=0,1,...,m$ are defined as
\begin{equation}\label{adjsn.}
    \bigl(A_{k})_{i, j}\;=\left\{\begin{array}{c}
      \hspace{-2.65cm}1 \quad \mathrm{if} \;\;\   \partial(i,j)=k, \\
      0 \quad \quad \mathrm{otherwise} \quad \quad \quad  (i, j
    \in M^{\lambda}) \\
    \end{array}\right. , \;\;\ k=0,1,...,m.
\end{equation}
where, we mean by $\partial(i,j)$ the number of components that
$i=(i_1,...,i_m)$ and $j=(j_1,...,j_m)$ are different (this is the
same as Hamming distance which is defined in coding theory). The
network with adjacency matrices defined by (\ref{adjsn.}) is known
also as the Johnson network $J(n,m)$ and has $m+1$ strata such
that
\begin{equation}\label{stratsn}
\kappa_0=1,\;\ \kappa_l=\left(\begin{array}{c}
                            m \\
                            m-l
                          \end{array}\right)\left(\begin{array}{c}
                            n-m \\
                            l
                          \end{array}\right) ,\;\;\ l=1,2,...,m.
\end{equation}
One should notice that for the purpose of PST, we must have
$\kappa_m=1$ which is fulfilled if $n=2m$, so we will consider the
network $J(2m,m)$ in order to transfer $m$-qubit $GHZ$ states
(hereafter we will take $n=2m$ so that we have $\kappa_m=1$). If
we stratify the network $J(2m,m)$ with respect to a given
reference node $\ket{\phi_0}=\ket{i_1,i_2,...,i_m}$, the unit
vectors $\ket{\phi_i}$, $i=1,...,m$ are defined as
$$\ket{\phi_1}=\frac{1}{\sqrt{\kappa_1}}\sum_{i'_1\neq i_1}\ket{i'_1,i_2,...,i_m}+\sum_{i'_2\neq i_2}\ket{i_1,i'_2,i_3,...,i_m}+...+\sum_{i'_m\neq i_m}\ket{i_1,...,i_{m-1},i'_m},$$
$$\ket{\phi_2}=\frac{1}{\sqrt{\kappa_2}}\sum_{k\neq l=1}^m\sum_{i'_l\neq i_l,i'_k\neq i_k}\ket{i_1,...i_{l-1},i'_l,i_{l+1},...,i_{k-1},i'_k,i_{k+1}...,i_m},$$
\vdots
$$\ket{\phi_j}=\frac{1}{\sqrt{\kappa_j}}\sum_{k_1\neq k_2\neq...\neq k_j=1}^m\sum_{i'_{k_1}\neq i_{k_1},...,i'_{k_j}\neq i_{k_j}}\ket{i_1,...,i_{k_1-1},i'_{k_1},i_{k_1+1},...,i_{k-1},i'_{k_j},i_{k_j+1}...,i_m},$$
\vdots
\begin{equation}\label{unitvsn}
\ket{\phi_m}=\frac{1}{\sqrt{\kappa_m}}\sum_{i'_{1}\neq
i_{1},...,i'_{m}\neq i_{m}}\ket{i'_1,i'_2,...,i'_{m}}.
\end{equation} Since the network $J(n,m)$ is distance-regular, the above stratification is independent of the choice of
reference node. The intersection array of the network is given by
\begin{equation}\label{intsn}
b_l=(m-l)^2\;\  ;  \;\;\ c_l=l^2.
\end{equation}
Then, by using the Eq. (\ref{omegal}), the QD parameters
$\alpha_i$ and $\omega_i$ are obtained as follows
\begin{equation}\label{QDsn}
\alpha_l=2l(m-l) \;\ l=0,1,...,m  ;  \;\;\
\omega_l=l^2(m-l+1)^2\;\ , l=1,2,...,m.
\end{equation}
Then, one can show that \cite{jss1}
\begin{equation}\label{QDRsn}
A\ket{\phi_l}=(l+1)(m-l)\ket{\phi_{l+1}}+2l(m-l)\ket{\phi_l}+l(m-l+1)\ket{\phi_{l-1}}.
\end{equation}
\section{Perfect Transfer of $m$-qubit GHZ states over antipodes of the network $J(2m,m)$}
The quantum state transfer protocol involves two steps:
initialization and evolution. First, a $m$-qubit $GHZ$ state
$\ket{\psi}_A=\frac{1}{\sqrt{2}}(\ket{00\ldots 0}_A+\ket{11\ldots
1}_A)\in {\mathcal{H}}_A$ to be transmitted is created. The state
of the entire system after this step is given by
\be\label{eq1}\ket{\psi(t=0)}=\ket{\psi}_A\ket{\underbrace{00\ldots0}_m}_B=\frac{1}{\sqrt{2}}(\ket{00\ldots0}_A\ket{00\ldots0}_B+\ket{11\ldots1}_A\ket{00\ldots0_B}).\ee
Then, the network couplings are switched on and the whole system
is allowed to evolve under $U(t)=e^{-iHt}$ for a fixed time
interval, say $t_0$. The final state becomes \be \ket{\psi(t_0)}
=\frac{1}{\sqrt{2}}(\ket{00\ldots0}_A\ket{00\ldots0}_B+\sum_{j=1}^{2m}f_{jA}(t_0)\ket{j})
\ee where, $\ket{j}\equiv \ket{j_1j_2\ldots
j_m}=\ket{0\ldots0\underbrace{1}_{j_1}0\ldots0\underbrace{1}_{j_2}0\ldots0\underbrace{1}_{j_m}0}$
and
$\ket{A}=\ket{\underbrace{11\ldots1}_m\underbrace{00\ldots0}_m}$
so that $f_{jA}(t_0):=\langle j|e^{-iHt_0}|A\rangle$. Any site $B$
is in a mixed state if $|f_{AB}(t_0)|<1$, which also implies that
the state transfer from site $A$ to $B$ is imperfect. In this
paper, we will focus only on PST. This means that we consider the
condition \be\label{eq3} |f_{AB}(t_0)|=1\;\;\ \mbox{for}\;\
\mbox{some}\;\ 0<t_0<\infty\ee
 which can be interpreted as the signature of
perfect communication (or PST) between $A$ and $B$ in time $t_0$
(with $\ket{B}=\ket{00\ldots011\ldots1}$). The effect of the
modulus in (\ref{eq3}) is that the state at $B$, after
transmission, will no longer be $\ket{\psi}$, but will be of the
form \be\frac{1}{\sqrt{2}}(\ket{00\ldots0}+e^{i\phi}\ket{11\ldots
1}). \ee The phase factor $e^{i\phi}$ is not a problem because
$\phi$ can be corrected for with an appropriate phase gate (for
more details see for example \cite{3'',9'',yung,yung1}).

As regards the arguments of subsection $2.2$, the evolution with
the adjacency matrix $H=A\equiv A_1$ for distance-regular networks
starting in $\ket{\phi_0}$, always remains in the stratification
space. For distance-regular network $J(2m,m)$ for which the last
stratum, i.e., $\ket{\phi_{m}}$ contains only one site, then PST
between the antipodes $\ket{\phi_0}$ and $\ket{\phi_{m}}$ is
allowed. Thus, we can restrict our attention to the stratification
space for the purpose of PST from  $\ket{\phi_0}$ to
$\ket{\phi_{m}}$.

The model we will consider is the network $J(2m,m)$ consisting of
$N=C^{2m}_m=\frac{(2m)!}{m!m!}$ sites labeled by $\{1,2, ... ,N\}$
and diameter $m$. Then we stratify the network with respect to a
chosen reference site, say $1$. At time $t=0$, the state in the
first (input) site of the network is prepared in the $m$-qubit
$GHZ$ state $\ket{\psi_{in}}$. We wish to transfer the state to
the $N$th (output) site of the network with unit efficiency after
a well-defined period of time. We shall assume that initially  the
network is in the state $\ket{00\ldots0}_A\ket{00\ldots0}_B$.
Then, we consider the dynamics of the system to be governed by the
quantum-mechanical Hamiltonian
\begin{equation}\label{H}
H_G =\sum_{k=1}^mJ_{k}P_k(1/2\sum_{_{1\leq i<j\leq
2m}}{\mathbf{\sigma}}_i\cdot {\mathbf{\sigma}}_j+\frac{m}{2}I),
\end{equation}
where, ${\mathbf{\sigma}}_i$ is a vector with familiar Pauli
matrices $\sigma^x_i, \sigma^y_i$ and $\sigma^z_i$  as its
components acting on the one-site Hilbert space ${\mathcal{H}}_i$,
and $J_{m}$ is the coupling strength between the reference site
$1$ and all of the sites belonging to the $k$-th stratum with
respect to $1$.

In order to use the spectral analysis methods, we write the
hamiltonian (\ref{H}) in terms of the adjacency matrix $A$ of the
network $J(2m,m)$. To do so, first we note that for the Johnson
network $J(n,m)$ with adjacency matrix $A$ one can show that
\be\label{perm} \sum_{1\leq i<j\leq
n}P_{ij}=A+[\left(\begin{array}{c}
                             m \\
                             2\\
                                                \end{array}\right)+\left(\begin{array}{c}
                             n-m \\
                             2\\
                                                \end{array}\right)
                                              ]I, \ee
where $P_{ij}$ is the permutation operator acting on sites $i$ and
$j$. In fact restriction of the operator $\sum_{1\leq i<j\leq
n}P_{ij}$ on the $m$-particle space (space spanned by the states
with $m$ spin down) which has dimension $C^n_m$, is written as the
adjacency matrix $A$ of the network as in the Eq. (\ref{perm}).
For $n=2m$, the Eq.(\ref{perm}) is written as \be\label{perm1}
\sum_{1\leq i<j\leq 2m}P_{ij}=A+m(m-1)I. \ee Then, by using the
fact that
$$\sigma_i\cdot \sigma_j=2P_{ij}-I \;\ \rightarrow \;\ \frac{1}{2}\sum_{_{1\leq i<j\leq
2m}}{\mathbf{\sigma}}_i\cdot {\mathbf{\sigma}}_j=\sum_{_{1\leq
i<j\leq 2m}}P_{ij}-\frac{1}{2}\left(\begin{array}{c}
                             2m \\
                             2\\
                                                \end{array}\right)I=A-\frac{m}{2}I,$$ we
obtain
\begin{equation}\label{H1}
H_G =\sum_{k=1}^mJ_{k}P_k(A).
\end{equation}

Now, for the purpose of the perfect transfer of a $m$- qubit GHZ
state, we impose the constraints that the amplitudes
$\langle\phi_i|e^{-iHt}|\phi_0\rangle$ be zero for all
$i=0,1,...,m-1$ and
$\langle\phi_m|e^{-iHt}|\phi_0\rangle=e^{i\theta}$, where $\theta$
is an arbitrary phase. Therefore, these amplitudes must be
evaluated. To do so, we use the stratification and spectral
distribution associated with the networks $J(2m,m)$ to write
$$\langle\phi_i|e^{-iHt}|\phi_0\rangle=\langle\phi_i|e^{-it\sum_{l=0}^mJ_lP_l(A)}|\phi_0\rangle=\frac{1}{\sqrt{\kappa_i}}\langle\phi_0|A_ie^{-it\sum_{l=0}^mJ_lP_l(A)}|\phi_0\rangle$$
Let the spectral distribution of the graph is
$\mu(x)=\sum_{k=0}^m\gamma_k\delta(x-x_k)$ (see Eq. (\ref{m})).
The Johnson network is a kind of network with a highly regular
structure that has a nice algebraic description; For example, the
eigenvalues of this network can be computed exactly (see for
example the notes by Chris Godsil on association schemes
\cite{Godsil} for the details of this calculation). Indeed, the
eigenvalues of the adjacency matrix of the network $J(2m,m)$ (that
is $x_k$'s in $\mu(x)$) are given by \be\label{eig}
x_k=m^2-k(2m+1-k),\;\;\ k=0,1,\ldots ,m.\ee Now, from the fact
that for distance-regular graphs we have
$A_i=\sqrt{\kappa_i}P_i(A)$ \cite{jss1},
$\langle\phi_i|e^{-iHt}|\phi_0\rangle=0$ implies that
$$\sum_{k=0}^m\gamma_kP_i(x_k)e^{-it\sum_{l=0}^mJ_lP_l(x_k)}=0, \;\;\ i=0,1,...,m-1$$
Denoting $e^{-it\sum_{l=0}^mJ_lP_l(x_k)}$ by $\eta_k$, the above
constraints are rewritten as follows
$$
\sum_{k=0}^mP_i(x_k)\eta_k\gamma_k=0,\;\;\ i=0,1,...,m-1,$$
\begin{equation}\label{Cons.}
\sum_{k=0}^mP_m(x_k)\eta_k\gamma_k=e^{i\theta}.
\end{equation}
From invertibility of the matrix ${\mathrm{P}}_{ik}=P_i(x_k)$ (see
Ref. \cite{pstd}) one can rewrite the Eq. (\ref{Cons.}) as
\begin{equation}\label{Cons.1}
\left(\begin{array}{c}
  \eta_0\gamma_0 \\
  \eta_1\gamma_1 \\
  \vdots \\
  \eta_d\gamma_d
\end{array}\right)=P^{-1}\left(\begin{array}{c}
  0 \\
  \vdots\\
  0 \\
  e^{i\theta}
\end{array}\right).
\end{equation}
The above equation implies that $\eta_k\gamma_k$ for $k=0,1,...,m$
are the same as the entries in the last column of the matrix
${\mathrm{P}}^{-1}=WP^t$ multiplied with the phase $e^{i\theta}$,
i.e., for the purpose of PST, the following equations must be
satisfied \be\label{result}
\eta_k\gamma_k=\gamma_ke^{-2it_0\sum_{l=0}^mJ_lP_l(x_k)}=e^{i\theta}{(W{\mathrm{P}}^{t})}_{km}\;\
, \;\;\ \mbox{for} \;\ k=0,1,...,m ,\ee with
$W:=diag(\gamma_0,\gamma_1,\ldots,\gamma_{m})$.

One should notice that, the Eq.(\ref{result}) can be rewritten as
\be\label{res'}
(J_0,J_1,\ldots,J_D)=-\frac{1}{2t_0}[\theta+(2l_0+f(0))\pi,\theta+(2l_1+f(1))\pi,\ldots,\theta+(2l_D+f(D))\pi](W{\mathrm{P}}^{t}),\ee
or \be\label{res''}
J_k=-\frac{1}{2t_0}\sum_{j=0}^m[\theta+(2l_j+f(j))\pi](W{\mathrm{P}}^{t})_{jk},\ee
where $l_k$ for $k=0,1,\ldots, m$ are integers and $f(k)$ is equal
to $0$ or $1$ (we have used the fact that $\gamma_k$ and
${(W{\mathrm{P}}^{t})}_{km}$ are real for $k=0,1,\ldots, m$, and
so we have $\gamma_k=|{(W{\mathrm{P}}^{t})}_{km}|$). The result
(\ref{res''}) gives an explicit formula for suitable coupling
constants so that PST between the first node ($\ket{\phi_0}$) and
the opposite one ($\ket{\phi_m}$) can be achieved.

In the following we consider PST of the two qubit (the case $m=2$)
GHZ state $\ket{\psi}_A=\frac{1}{2}(\ket{00}+\ket{11})$ in
details: From Eq. (\ref{QDsn}), for $m=2$, the QD parameters are
given by
$$\alpha_1=2,\;\ \alpha_2=0; \;\;\ \omega_1=\omega_2=4,$$
Then by using the recursion relations (\ref{op}) and (\ref{oq}),
we obtain
$$Q_2^{(1)}(x)=x^2-2x-4,\;\;\ Q_3(x)=x(x-4)(x+2),$$
so that the stieltjes function is given by
$$G_{\mu}(x)=\frac{Q_2^{(1)}(x)}{Q_3(x)}=\frac{x^2-2x-4}{x(x-4)(x+2)}.$$
Then the corresponding spectral distribution is given by
$$\mu(x)=\sum_{l=0}^2\gamma_l\delta(x-x_l)=\frac{1}{6}\{3\delta(x)+\delta(x-4)+2\delta(x+2)\},$$
which indicates that
$$W=\left(\begin{array}{ccc}
          \gamma_0 &0 &0 \\
            0 & \gamma_1 &0 \\
            0& 0 & \gamma_2 \\
          \end{array}\right)=\frac{1}{6}\left(\begin{array}{ccc}
          3 &0 &0 \\
            0 & 1 &0 \\
            0& 0 & 2 \\
          \end{array}\right).$$
In order to obtain the suitable coupling constants, we need also
the eigenvalue matrix $P$ with entries
$P_{ij}=P_i(x_j)=\frac{1}{\sqrt{\omega_1\ldots
\omega_i}}Q_i(x_j)$. By using the recursion relations (\ref{op}),
one can obtain $P_0(x)=1,\;\ P_1(x)=\frac{x}{2}$ and
$P_2(x)=\frac{1}{4}(x^2-2x-4)$, so that
$$P=\left(\begin{array}{ccc}
          1 &1 & 1 \\
            0 & 2 &-1 \\
            -1& 1 & 1 \\
          \end{array}\right).$$ Now, from the result (\ref{res''}), we obtain
  $$J_0=-\frac{6\theta+6\pi}{12t_0},\;\ J_1=-\frac{\pi}{3t_0},\;\ J_2=\frac{\pi}{12t_0}.$$
\section{Conclusion}
By using spectral analysis methods and employing algebraic
structures of Johnson networks $J(2m,m)$ such as
distance-regularity and stratification, a method for finding a
suitable set of coupling constants in some particular spin
Hamiltonians associated with the networks was given so that PST of
$m$-qubit GHZ states between antipodes of the
networks can be achieved. \\


\begin{thebibliography}{99}
\bibitem{PST}M. A. Jafarizadeh and R. Sufiani, (2008), Phys. Rev. A 77,
022315.
\bibitem{pstd}M. A. Jafarizadeh, R. Sufiani, S. F. Taghavi and E. Barati (2008),
arXiv: quant-ph/0803.2334.
\bibitem{1'} D. P. DiVincenzo, D. Bacon, J. Kempe, G. Burkard, and K. B.
Whaley, Nature 408, 339 (2000).
\bibitem{3'} F. Verstraete, M. Popp, and J.
I. Cirac, Phys. Rev. Lett. 92, 027901 (2004)
\bibitem{10'} F. Verstraete,1 M. A. Martý-Delgado, and J. I. Cirac1Phys. Rev. Lett.
92, 087201-1(2004).
\bibitem{4'} C. P. Sun, Y. Li, and X. F. Liu, Phys.
Rev. Lett. 91, 147903 (2003).
\bibitem{Bose}
Sougato Bose, Phys. Rev. Lett. 91, 207901 (2003).
\bibitem{2''} V.
Subrahmanyam, Phys. Rev. A 69, 034304 (2004)
\bibitem{3''}
M. Christandl, N. Datta, A. Ekert and A. J. Landahl, Phys. Rev.
Lett. 92, 187902 (2004).
\bibitem{9''}
M. Christandl, N. Datta, T. C. Dorlas, A. Ekert, A. Kay and A. J.
Landahl, (2005), Phys. Rev. A 71, 032312.
\bibitem{4''} C. Albanese, M. Christandl, N. Datta
and A. Ekert, Phys. Rev. Lett. 93, 230502 (2004).
\bibitem{Os}
T. J. Osborne and N. Linden, Phys. Rev. A 69, 052315 (2004).
\bibitem{6''}
H. L. Haselgrove, quant-ph/0404152.
\bibitem{7''}
F. Verstraete, M. A. Martin-Delgado and J. I. Cirac, Phys. Rev.
Lett. 92, 087201 (2004).
\bibitem{8''}F. Verstraete, M. Popp and J. I.
Cirac, Phys. Rev. Lett. 92, 027901.
\bibitem{9''}B.Q. Jin and V. E. Korepin, Phys. Rev. A 69, 062314 (2004).
\bibitem{10''} M. H Yung, D. W Leung and S. Bose, Quant. Inf. and Comp. 4, 174
(2004).
\bibitem{11''}L. Amico, A. Osterloh, F. Plastina, R. Fazio and G.
M. Palma, Phys. Rev. A 69, 022304 (2004).
\bibitem{12''}V. Giovannetti and
R. Fazio, quant-ph/0405110.
\bibitem{Bu}
D. Burgarth and S. Bose, (2005), Phys. Rev. A 71, 052315.
\bibitem{Bu1}
D. Burgarth and S. Bose, (2005), New J. Phys. 7, 135.
\bibitem{spin1} A. Chan and R. Hosoya, (2004), J. Algebraic
Combinatorics 20, 341-351.
\bibitem{spin2} A. Chan and C. D. Godsil, (2004), J. Combin. Th.
Ser. A 106, 165-191.
\bibitem{13''}M.B. Plenio, J. Hartley and J.
Eisert, New J. of Phys. 6, 36 (2004).
\bibitem{krylov} M. A. Jafarizadeh, R. Sufiani, S. Salimi and S.
Jafarizadeh, Eur. Phys. J. B 59, 199-216.
\bibitem{js}
M. A. Jafarizadeh and S. Salimi,(2006), J. Phys. A : Math. Gen.
39, 1-29.
\bibitem{jss1}
M. A. Jafarizadeh, R. Sufiani and S. Jafarizadeh, (2007), J. Phys.
A: Math. Theor. 40, 4949-4972.
\bibitem{js1}
M. A. Jafarizadeh, S. Salimi, (2007), Annals of physics, Vol. 322
1005-1033.
\bibitem{jss} M. A. Jafarizadeh, R. Sufiani, S. Salimi
and S. Jafarizadeh, (2007), Eur. Phys. J. B 59, 199-216.
\bibitem{simon}
H. Cycon, R. Forese, W. Kirsch and B. Simon {\it Schrodinger
operators} (Springer-Verlag, 1987).
\bibitem{Hislop}
P. D. Hislop and I. M. Sigal, {\it Introduction to spectral
theory: With applications to schrodinger operators} (1995).
\bibitem{Kim}
Y. S. Kim, {\it Phase space picture of quantum mechanics:group
theoretical approach}, (Science, 1991).
\bibitem{Hai}
H. W. Lee, Physics. Report, \textbf{259}, 147 (1995).
\bibitem{obata}
N. Obata, (2004), Quantum Probabilistic Approach to Spectral
Analysis of Star Graphs, Interdisciplinary Information Sciences,
Vol. 10, 41-52.
\bibitem{obh}
A. Hora, and N. Obata, (2003), Fundamental Problems in Quantum
Physics, World Scientific, 284.
\bibitem{st}
J. A. Shohat, and J. D. Tamarkin, (1943), {\it The Problem of
Moments, American Mathematical Society}, Providence, RI.
\bibitem{tsc} T. S. Chihara, (1978), {\it An Introduction to Orthogonal Polynomials}, Gordon and
Breach, Science Publishers Inc.
\bibitem{obah}
A. Hora, and N. Obata, (2002), Quantum Information V, World
Scientific, Singapore.
\bibitem{yung} M. H. Yung and S. Bose, (2005), Phys. Rev. A 71,
032310.
\bibitem{yung1} M. H. Yung, (2006), Phys. Rev. A 74, 030303.
\bibitem{Godsil} C. Godsil, (2005), {\it Association schemes}, Combinatorics
and Optimization, University of Waterloo
\end{thebibliography}
\end{document}